\documentclass[usegraphicx,usenatbib]{mn2e}
\usepackage{times}

\usepackage[T1]{fontenc} 
\usepackage{aecompl}

\usepackage{amsmath,bm, array,amssymb,macros}
\usepackage[colorlinks]{hyperref}
\hypersetup{colorlinks=true,
            breaklinks=true,
            linkcolor=black,
            urlcolor=black,
            citecolor=black,
	    }
\newcommand{\ten}[1]{\times 10^{#1}}
\newcommand{\Msun}{\mathrm{M}_{\odot}}

\newcommand{\macs}{MACS J1149.5+2223~}
\newcommand{\macswo}{MACS J1149.5+2223}

\newcommand{\pp}{RVW2~}
\newcommand{\ppwo}{RVW2}
\newcommand{\asec}{\mathrm{~arcsec}}

\title[Lensing Model of MACS J1149.5+2223 - I. Cluster mass reconstruction]
{Lensing Model of MACS J1149.5+2223 - I. Cluster mass reconstruction}

\author[Rau et al.]
       { S. Rau		$^{1}$ \thanks{Email: rau@mpa-garching.mpg.de},
         S. Vegetti	$^{1}$,
         S. D. M. White	$^{1}$
\\
$^1$Max-Planck Institute for Astrophysics, Karl-Schwarzschild Str. 1, D-85748, Garching, Germany \\
}

\date{Accepted 2014 June 13. Received 2014 May 14; in original form 2014 February 28}
\volume{443} \pagerange{957--968} \pubyear{2014}

\begin{document}
\maketitle
\author{Stefan Rau, Simona Vegetti, Simon White}

\begin{abstract}
Measurements of the total logarithmic central slope of the mass profile in galaxy clusters constrain their evolution and 
assembly history and that of their brightest cluster galaxies.
We report the first full surface brightness distribution modelling of the inner region of the galaxy cluster MACS J1149.5+2223.
We compare these results with a position-based modelling approach for which we employ more than twice the previously known positional constraints.
This is the first time that the detailed lensed image configuration of two non-central cluster galaxies with Einstein rings has been mapped.
Due to the extended radial coverage provided by the multiple images in this system,
we are able to determine the slope $\partial \log{ \kappa  }/\partial \log{R} = -0.33$ of the total projected mass distribution
from $8$ to $80~\mathrm{kpc}$. This is within the cluster-to-cluster scatter estimates from previous cluster measurements.
Our reconstruction of the image surface brightness distribution of the large central spiral galaxy has a root mean square residual for all image pixels
of $1.14~\sigma$, where $\sigma$ is the observational background noise. This corresponds to a reconstruction of the positions of bright clumps 
in the central galaxy with an rms of $0.063~\mathrm{arcsec}$.
\end{abstract}


\section{Introduction}
Our current understanding of structure formation is based on a hierarchical picture where more massive dark matter (DM)
structures form via the infall and the progressive merging of smaller objects. 
DM-only simulations, within the cold dark matter (CDM) paradigm, have predicted, over the last two decades,
remarkably self-similar DM mass profiles that are well fitted by 
gradually steepening analytic models, such as the Einasto \citep{Einasto_Profile} or the
Navarro, Frenk and White \citep[NFW;][]{NFW_Profile} profile, over a wide range of scales from galaxy clusters
\citep{Springel_cluster_2001, Diemand_cluster_2004, Reed_cluster_2005,Springel_nature_Millenium_2005,Merritt_2006,Gao_Phoenix}
to galaxies \citep{Navarro_Aquarius_galaxy_2010}. A central slope $\gamma_1 = \partial \log{\rho} / \partial \log{r} \approx  -1$ and a 
steeper outer slope $\gamma_2 \approx -3$ are generally found in those simulations. However, the details of the mass distribution are still being refined, and in general, 
there is no definitive prediction from numerical simulations on the asymptotic central slope of the galaxy cluster mass distribution.
In a suite of nine high-resolution simulated N-body DM-only clusters, \cite{Gao_Phoenix} have found considerable ($\sim 20$ percent) halo-to-halo scatter.	
% At the galaxy scale, environmental and baryonic effects become increasingly important
% in shaping the DM mass distribution. 
On scales of a few $\mathrm{kpc}$, the baryonic component of the brightest cluster galaxy (BCG) and the dark matter halo play comparable roles in shaping the total density.
However, there are competing hypotheses about their relative importance and the relevant physical processes involved.
The inner DM profiles can be steepened by adiabatic contraction
\citep{Gnedin_2004, Sellwood_2005, Sommer-Larsen_2010},
or flattened by (repeated) ejection of gas by active galactic nuclei (AGN) \citep{Martizzi_2013}
or by heating by dynamical friction \citep{ElZant_2001,ElZant_2004}.
Even in simulations with dissipationless mergers of multi-component systems (DM+stars), a flattening of the inner DM profile can be achieved \citep{Laporte_2012}.

Gravitational lensing provides a powerful tool to compare the predictions of cluster simulations with observations. 
There has been a considerable effort to get a coherent picture of the total cluster mass distribution on a large range of scales 
by combining stellar kinematics at the innermost radii ($r<10~\mathrm{kpc}$) with weak lensing measurements at radii $r>100~\mathrm{kpc}$ and strong lensing 
on intermediate scales \citep[e.g.][]{Sand_2004,Sand_2008,Newman_2013a}.
 In particular, \cite{Sand_2008} combined stellar velocity dispersion observations with strong-lensing data for the cluster Abell 383 and
found that the inner slope of the dark matter is $\gamma_{\mathrm{DM}} \approx -0.45$. This is shallower than the values found at similar radii in purely DM simulations.
More recent observational constraints come from the analysis by \citet{Newman_2013a}. 
These authors have performed a combined stellar kinematic, strong lensing and weak lensing mass reconstruction 
for seven massive $(0.4 < M_{200}/ (10^{15} \Msun) <2)$\footnote{$M_{200}$ is the mass within a sphere with radius $r_{200}$ and a mean enclosed density 200 times the critical value}
clusters with redshifts between 0.2 and 0.3. 
They find a 3D central logarithmic total mass density slope of $\gamma \approx -1.16$ consistent with the NFW profile and with a scatter 
between clusters of $\sigma_\gamma = 0.13 \left( 68 \% \mathrm{CL}\right)$. 
This, however, is the total density profile, and the central $<10~\mathrm{kpc}$ are typically dominated by the stars of the BCG. 
An analysis of the separate contributions of baryons and DM in \citet{Newman_2013b} indicates a flattening of the DM profile
with a logarithmic central slope of $\gamma_{\mathrm{DM}} \approx -0.5$.

In this paper, we focus on building a high resolution lens model for \macswo~and on comparing the slope inferred from the lensing reconstruction with the predictions
for the total central mass density slope of galaxy clusters from cosmological structure formation simulations.

The cluster \macswo~was initially discovered as part of the Massive Cluster Survey (MACS) quest for the most massive galaxy clusters in the Universe \citep{Ebeling01,Ebeling07}.
More recently, \citet{Zheng_2012_magnified} have used the magnifying power of this cluster to identify a faint galaxy located at redshift $z=9.6\pm0.2$.

So far, two gravitational lens models for the cluster \macs have been published.
The first model, by \cite{Zitrin_2009}, assumes that the mass approximately follows the light.
The model consists of a superposition of power-law mass profiles for each galaxy in the cluster. 
As constraints, this model uses the multiple image positions from strong lensing. However, many of the details, especially 
for system 1 (see Sec.~\ref{sec:observations_and_constraints}), are only reproduced approximately. 
Under the assumption that the BCG consists only of stellar mass, this models infers a nearly uniform DM surface mass density out to $\sim 200~\mathrm{kpc}$.

The second model, by \cite{Smith_2009}, is also based on the image positions of multiply lensed bright clumps. 
However,  it makes us of a larger number of bright clumps, hence a significantly larger number of constraints.
The reconstructed image positions have a root mean square (rms) deviation from the observed positions of $0.5 \asec$.
This model rules out the flat central profile proposed by \cite{Zitrin_2009}.

In this work, we improve on these previous models in three aspects. 
First, we use a more sophisticated model. 
We model {\em all five} galaxies that are close to multiply lensed images and close to the cluster centre using
individual mass profiles. This is crucial for reproducing the morphology of the lensed image of the main system (see Sec.~\ref{sec:observations_and_constraints}).
Secondly, we identify twice as many positional constraints as previously used. These include multiply-imaged clumps that are part of two Einstein rings formed by two cluster galaxies, as well as
details of the nonlinear configuration of the image covering the centre of the cluster. Thanks to the increased number of constraints, we can place tight constraints on the slope of the total mass density profile.
Thirdly, we use a more sophisticated gravitational lens modelling method that was originally developed and applied on galaxy scales by \citet{Koopmans_2005, Suyu_2006,Vegetti_2009}
and recently also on cluster scales by \cite{Eichner_2013_macsj1206}. In this method, we use the information provided by the positions of multiply-lensed images along-side
with the full surface brightness distribution of the images. 

This paper is organized as follows. We describe the observations, the image morphology of the main lensed image 
and the newly identified positional constraints in section \ref{sec:observations_and_constraints}.
In section \ref{sec:mass_model} we describe our analytic mass parametrization and in section \ref{sec:modelling_methods} our two modelling methods. 
Section \ref{sec:results} contains the main results, the best lens models for the position modelling and the surface brightness modelling in sections 
\ref{sec:results_modelling_pos} and \ref{sec:results_modelling_brightness} respectively. The slope measurement of the total mass distribution is presented in section \ref{sec:Central Slope}.
Throughout this paper, we assume $H_0 = 67.3$ and $\Omega_{\mathrm{m}} = 0.315$ from Planck \citep{Planck_Cosmology}. At the redshift of the cluster,
$z = 0.544$, 1 arcsec corresponds to $~6.6$~kpc. 

\section{Observations and Constraints}
\label{sec:observations_and_constraints}
\begin{figure*}
\begin{center}
\includegraphics[width=1.0\textwidth]{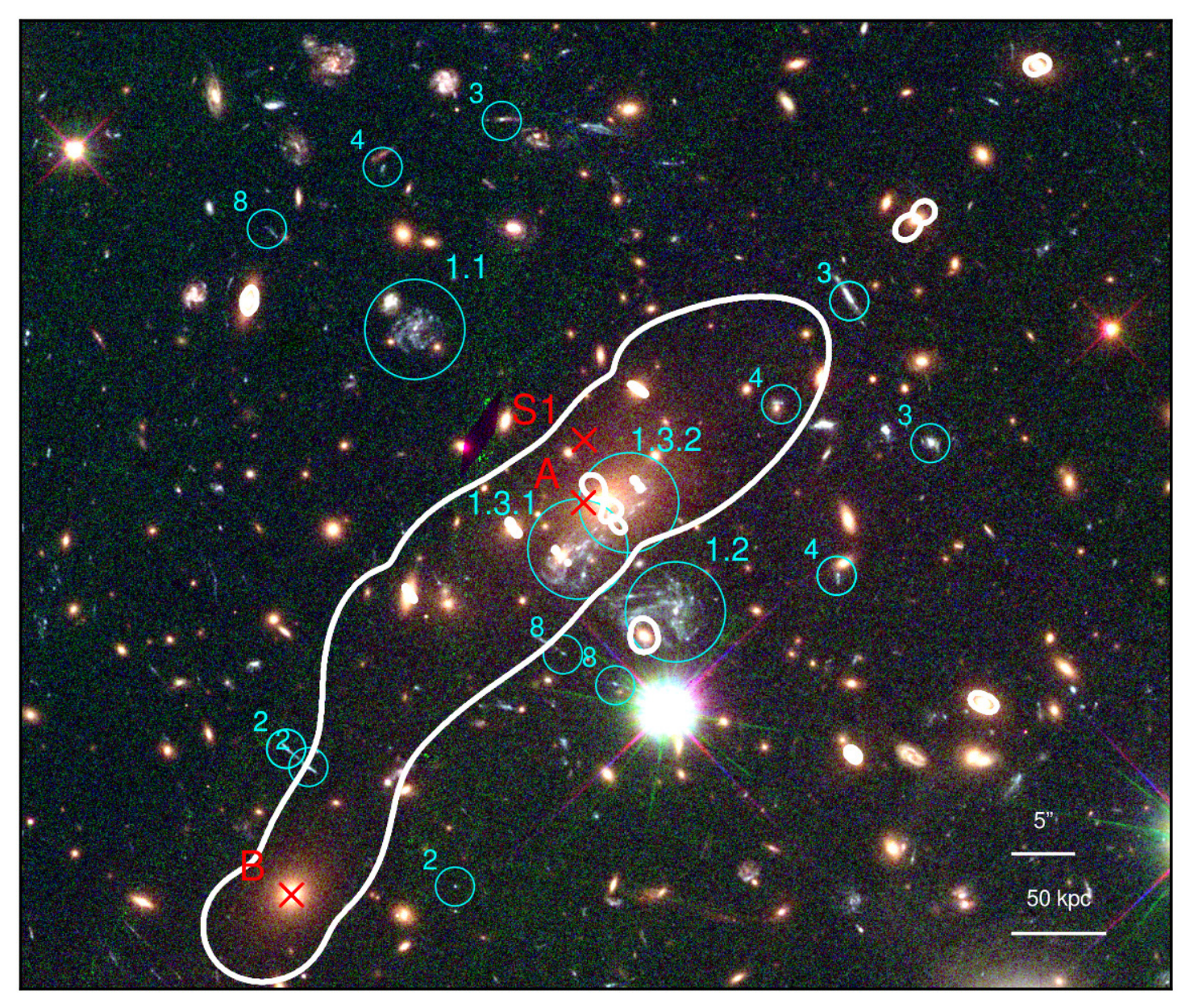}
\end{center}
\caption[Hubble Space Telescope observation of \macs]{
HST F814W/F606W/F555W RGB colour image from the Cluster Lensing And Supernova survey with Hubble (CLASH) observations of \macswo, (north is up and east is left).
Overlaid in white is the critical curve from our best model, for a source redshift of $z_s = 1.49$ and a cluster redshift of $z_l = 0.544$.
The centre of the reconstructed DM halo A is $\approx 1.5 \asec$ left of the BCG. 
There are three separate main images labelled 1. 
The detailed morphology of the central image, 1.3, is more complicated, parts have a seven fold image configuration.
Additional multiple images are labelled 2-8, for details see Sec.~\ref{sec:observations_and_constraints}.
\label{fig:HST_cluster}
}
\end{figure*}
The galaxy cluster \macs at redshift $z=0.54$ was observed as part of the Cluster Lensing And Supernova survey with Hubble (CLASH) programme 
with the Advanced Camera for Surveys (ACS) on board the  Hubble Space Telescope (HST). 
For details on the observations, the available filters and the imaging pipeline, we refer the reader to \cite{CLASH_overview_2012_short}.
The CLASH image pipeline provides redshifts of all galaxies close to the cluster. Spectroscopic redshifts are available for the sources 1, 2 and 3 from \cite{Smith_2009}.

Since our lens modelling technique makes use of both the positions of the lensed images and their surface brightness distribution, 
it is very important to minimize the light contamination from the lensing galaxies.
This is particularly true for those images that lie close to bright cluster members.
In this paper, we focus on the lens modelling of the F555W filter of the ACS, this being a compromise between a high signal-to-noise ratio and 
weak light contamination from the cluster galaxies in a single filter.	
We model the surface brightness distribution of the seven galaxies, the BCG and a star close to the three main lensed images, using the publicly available program {\sc GALFIT} \citep{galfit_2002}. 
The positions of all modelled components are summarized in Table~\ref{tab:brightness_subtraction}.
All light distributions are modelled as a Sersic profile. 

For the modelling based on the image positions only (see Sec.~\ref{sec:modelling_pos}), the robust and correct identification of many corresponding multiply lensed source clumps is crucial. 
Note, that instead the surface brightness reconstruction in Sec.~\ref{sec:modelling_hybrid} does {\em not} depend on somewhat arbitrarily chosen points on the image plane
since it uses the full image surface brightness distribution of the lensed images.
We identify five different sources lensed by the cluster \macs in as many as 15 images, 
here we count the main image system with a multiplicity of three, according to the brightest source clump. 
However, some bright clumps in source 1 are lensed up to seven times.
Figure~\ref{fig:HST_cluster} shows an overview of these sets of images. We follow the nomenclature of \cite{Smith_2009} and \cite{Zitrin_2009} to label the multiple images.
Spectroscopic observations on the Keck telescope in 2004 by \cite{Smith_2009} have found a redshift of $z=1.4906\pm0.0002$ for the source of image system 1, while the
sources 2 and 3 are located at redshifts $1.894$ and $2.497$, respectively.  
For sources 4 and 8 we take the photometric redshifts provided by the CLASH 16 band image pipeline of  3.0 and 2.9, respectively.

\begin{figure*}
\begin{center}
\includegraphics[width=2.0\columnwidth]{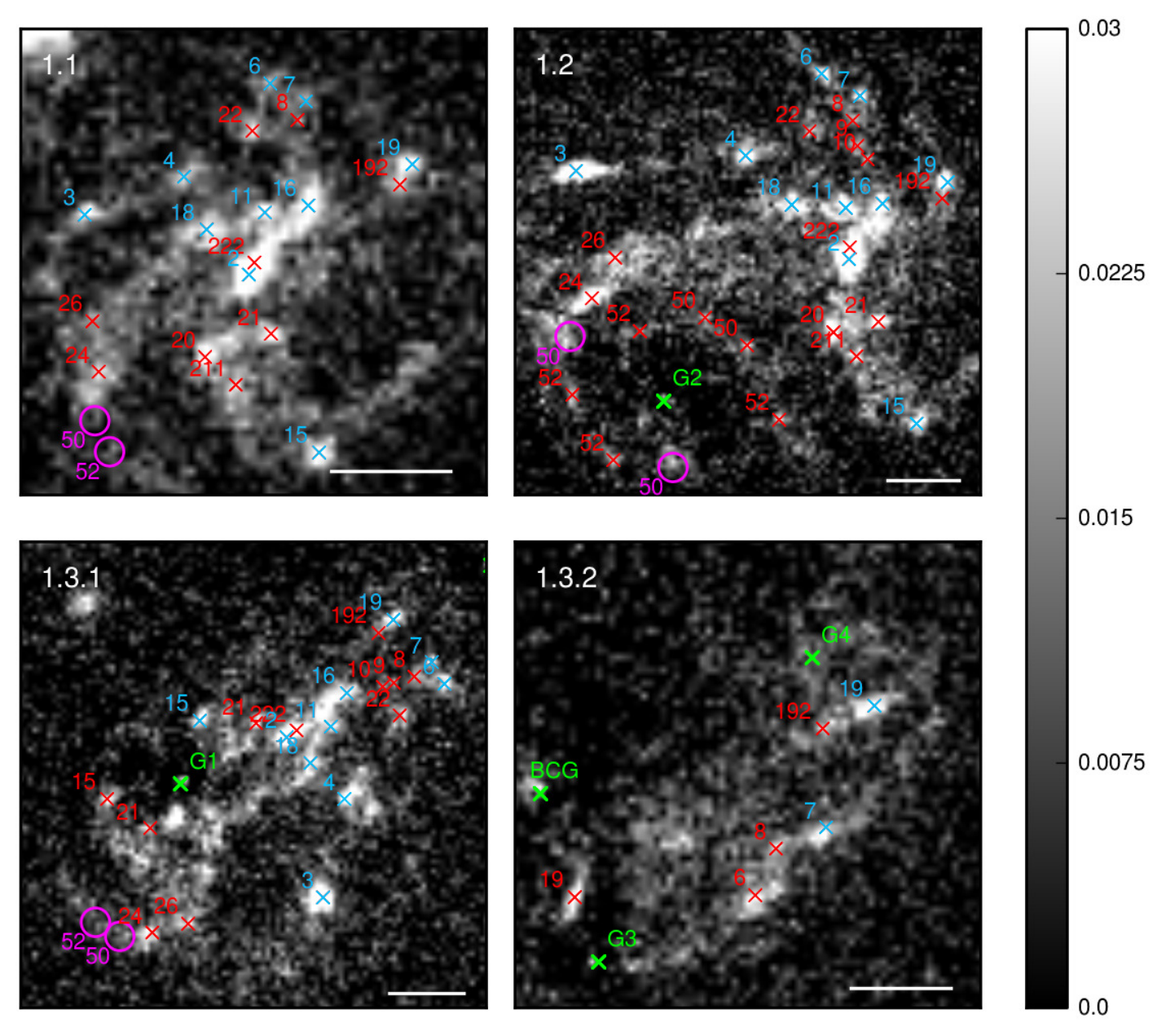}
\end{center}
\caption[Constraints for the image position modelling]{
Constraints used for the image position modelling. Top left: relatively undistorted source image. Top right: second image with a distinct signature of an additional
Einstein ring, around the satellite galaxy G2 (galaxy surface brightness subtracted). Bottom row: strongly distorted image system 3 close to the central galaxies. 
Green crosses indicate galaxies G1, G2, G3, G4 and the BCG, whose surface brightness distribution were subtracted in preparation for the hybrid modelling. 
Blue constraints are from \cite{Smith_2009}, red constraints are newly identified in this work. White line at the bottom indicates the $1\asec$ scale.
\label{fig:constraints_four}
}
\end{figure*}

Image system 1 consists of at least three lensed images of a spatially resolved spiral background galaxy and
it is currently one of the largest known gravitationally lensed image systems.
For a detailed view of each of the three main images 1.1, 1.2 and 1.3, we refer to Fig.~\ref{fig:constraints_four}.
We show the constraints identified by \cite{Smith_2009} in blue, and the multiply imaged source knots that we additionally identify and use in this work in red. 
Multiple images predicted by our best lens model for constraints 50 and 52 are indicated in magenta, those are {\em not} used as constraints in our modelling.

The farthest of the three main images is $1.1$, located at $(-16.2 \asec, 14 \asec)$
thus about $20 \asec$ from the BCG.
There are no massive cluster member galaxies nearby and the two smaller
close-by galaxies have no measurable lensing effect on the image surface brightness distribution. 
This image is roughly uniformly magnified and therefore gives a good impression of the almost unlensed source morphology.
For the first time, we are able to use constraints based on the lower left part of the source, 
consisting of the triply-imaged clumps $26$ and $24$ (cf. Fig.~\ref{fig:constraints_four}).

A second image $1.2$ is lensed at $(5.1 \asec,-9.1 \asec)$, a distance of $~11$ arcsec. 
If we compare the surface brightness distribution of images 1.2 and 1.1 in  Fig.~\ref{fig:constraints_four}, 
it is evident that the lower left spiral arm of the source in image 1.2 is additionally lensed. Indeed, galaxy G2 acts as an additional strong lens 
and deflects the two source clumps 50 and 52 into a \emph{secondary Einstein ring} with Einstein radius of about $1.3 \asec$. 
Due to the additional deflection by the galaxy G2, clump $52$ is lensed a total of seven times. 
Since the fifth image in image $1.2$ is lensed very close to the subtracted lens galaxy surface brightness distribution  
and since the corresponding multiple images in the main images $1.1$ and $1.3$ are not uniquely identifiable, 
we use, as constraints, only four of the seven multiple images. The additional magnification provided by the galaxy 
G2 significantly increases the visibility of the source clump $52$ in image $1.2$.
Identifying the location of multiple images by eye is sometimes
complicated and to some extent arbitrary. When trying to identify the multiple images of clump 50, we could not
easily and uniquely find more than two images. However,  from the inferred lens model we identify a posteriori 
additional multiple images in image 1.1, 1.2 and 1.3. (see the magenta circles in Figure \ref{fig:constraints_four}). 
This clearly shows the intrinsic limitations of position based modelling and the
necessity for a more complex approach that takes into account for the full surface brightness distribution.

The third lensed image, $1.3$, is lensed at a distance of $4.5 \asec$ from the BCG
at $(-2.9\asec,-4\asec)$ and has a significantly more complicated morphology. 
In Fig.~\ref{fig:constraints_four} we split this central image in two.
Image 1.3.1 is a full multiple image of the source galaxy. Note the cluster galaxy G1 that lies on top of the lensed image. 
This galaxy is responsible for additional strong lensing in image $1.3$ similar to galaxy G2 in image $1.2$. 
The newly identified clumps 15 and 21 are additional multiple images deflected by galaxy G1. 
Together with the positions of the new constraints 24 and 26, the mass distribution of the galaxy G1 is now tightly constrained
over a large range of radii from $r =0.4\asec$ $(2.6~\mathrm{kpc})$ from the galaxy centre to $r = 11\asec$ $(73~\mathrm{kpc})$.
Indeed, in the radial range covered by image 1.3, it is evident from the Einstein ring image configuration that galaxy G1
plays a dominant role. Beyond this distance, galaxy G1 is no longer the dominant deflector. However, we find, after having taken 
into account for the full degeneracies between the several mass components, that G1 still has a significant 
contribution of $>5$ percent to the total magnification at the distance of image 1.2 at $~11\asec$.

Finally, the most important and critical image is image 1.3.2. 
All multiple source clumps in this image are in the high magnification region
close to the cluster centre where there are three galaxies, the BCG, and the galaxies G3 and G4.
Image 1.3.2 consists of multiple images of only a few source clumps and was previously not constrained well. 
Here, we add the multiple source clumps 192, 8 and 6 as well as the fifth image of clump $19$ between images $1.3.1$ and $1.3.2$ to our list of constraints.
In total, we identify as many as 77 image positions in system 1 that we use to constrain the central regions of the cluster mass distribution.

At large scales the mass distribution of the cluster is constrained by the triple image systems 2, 3, 4 and 8. 
In total, these \emph{outer} constraints add up to 12 images rising from lensed sources located at redshifts spanning from $z = 1.5$ to $3.0$. 
The lower images of systems 4 and 8 provide constraints at distances between $13$ and $19~\mathrm{arcsec}$. 
Most constraints of systems 2, 3, 4 and 8, however, are in the range $26-33\asec$ from the BCG. 

\begin{table}
\label{tab:brightness_subtraction}
\caption[Positions of the surface brightness subtracted galaxies]
{Positions relative to the BCG of those galaxies for which the surface brightness distribution was modelled with a Sersic profile and then subtracted for the 
hybrid modelling.}
  \begin{center}
    \begin{tabular}{ c  c  c}\hline
  &  $x \left(\mathrm{arcsec}\right)$& $y \left(\mathrm{arcsec}\right)$ \\ \hline
BCG &  0&0 \\
G1 & -3.6&-4.6  \\
G2 & 3.2& -11.1 \\
G3 &  0.5&-1.6  \\ 
G4 & 2.6 & 1.4  \\ 
G5 &  -17.6&12.9 \\ 
G6 & -13.8&12.5 \\ 
G7 & -7.6&-2.3 \\ 
Star & 5.1&-17.2 \\ \hline
    \end{tabular}
  \end{center}
\end{table}

\section{The Mass Model}
\label{sec:mass_model}
\begin{figure*}
\begin{center}
% lbrt
% \includegraphics[trim=10mm 40mm 10mm 20mm,clip,width=2.0\columnwidth]{./Figure3.ps}
\includegraphics[trim=0mm 0mm 0mm 0mm,clip,width=2.0\columnwidth]{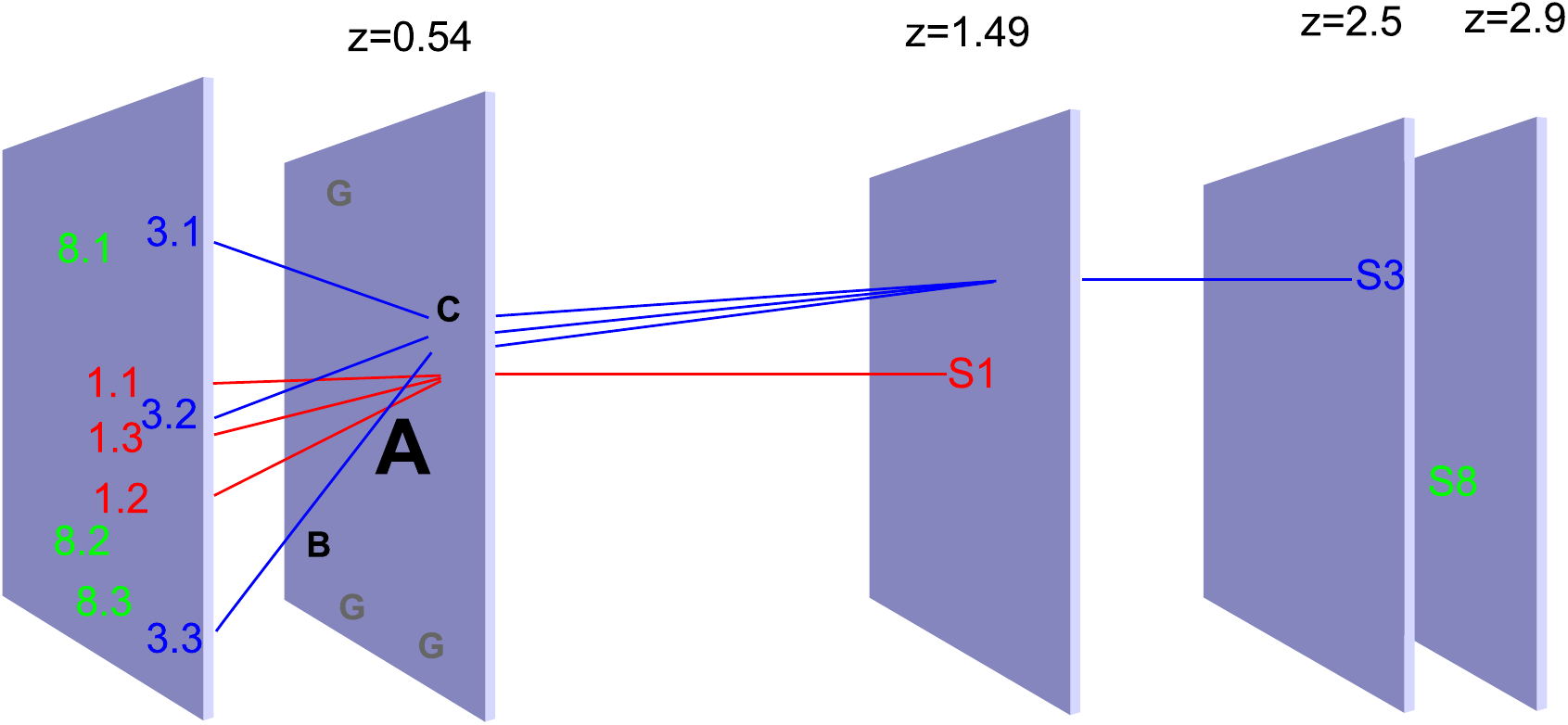}
\end{center}
\caption[Lensing Geometry]{
Geometry of the cluster lens and the sources S1, S3 and S8 and schematic light paths for sources S1 and S3 as an example. 
We include the lensing effect of the source S1 of the main image system 1,  in the reconstruction of all other multiple images whose sources are at higher redshift. 
A,B,C denote individually modelled mass components in the cluster, G scaled galaxy mass contributions.
\label{fig:Geometry}
}
\end{figure*}
In this section, we describe the analytical mass model and free parameters used for the lens modelling. 
The same model is used for both the image position modelling  (Sec.~\ref{sec:modelling_pos}) and the hybrid modelling of the image positions and the surface 
brightness distribution (Sec.~\ref{sec:modelling_hybrid}).

Following the CDM paradigm, the parametric mass model considered in this paper includes: a central dark matter halo for the cluster A, five central mass components for the BCG and the galaxies
G1, G2, G3 and G4, one mass component for a massive galaxy at $\mathrm{B} = (-25.7\asec, -32.3\asec)$, and one for a group of smaller galaxies at $\mathrm{C} = (19.2\asec, 48.1\asec)$ from the BCG. 
Our choice for the positions of A, B and C closely follows the ones chosen in \citet{Smith_2009} (A, B and D respectively, in their paper). We also stress that \citet{Smith_2009} 
did not explicitly include the galaxies G1 and G2 that will prove to be important in the detailed mass reconstruction of this cluster. 
Unlike \citet{Smith_2009}, we do not include the galaxy component E as well as the constraints around E located further north from C. 
This galaxy is located at a distance of $103\asec$ from the BCG that is roughly twice the distance to the mass component C. 
This choice is based on the fact that any contribution from a mass component at this distance would, at most, act as a uniform sheet of deflection 
with a vanishing gradient at the centre of the cluster. In this paper, we are interested in the central mass distribution and the total central slope of the cluster mass distribution. 
Due to the low amount of strong lensing constraints, the mass distribution of the cluster at these distances, $>30\asec$, can generally only be constrained by including weak lensing information.
Finally, our best reconstructed model has negligible shear in a direction not towards the component E (cf. Tab. \ref{tab:best_modelling_parameters}) indicating that the component E can be safely ignored.
Additionally, the normalization of the mass component C in our best model is relatively uncertain and therefore can absorb any small contribution from a mass component E within the errors of our best model. 
We also include one mass component for each of all the remaining identified cluster member galaxies in a scaled manner and a contribution from external shear.

Since the spiral source galaxy of image system 1, hereafter S1, lies at a redshift of $z = 1.49$, it has a significant lensing effect on the light ray paths coming from the other higher redshift sources.
We, therefore, include an extra mass component associated with S1 at $z = 1.49$ and employ a multiple lens plane algorithm. In particular, we use a nested loop reconstruction.
At the first step of the reconstruction we focus on the modelling of the image system 1, with S1 as background source and the full cluster as a lens. Since the source is an unknown of the lens modelling problem,
at this stage of the reconstruction, the position of S1 is an implicit free parameter of the model. At the next step we focus on the modelling of the image systems 2 to 8 by including the source S1 
as an additional lens. At this stage, the position of S1 is kept fixed at the previously inferred value, while the parameter describing its mass normalization is left free to vary.
Figure \ref{fig:Geometry} shows the geometry of the lens planes for sources 1, 3 and 8 and the schematic light paths for sources 1 and 3 as an example. 

We assume that all the above mass components have a total mass density distribution that follows a dual softened power-law elliptical mass distribution (dSPEMD) 
\citep{Barkana_fastell, Kassiola_1993_ellip_mass_models}, with projected surface mass density in units of the critical density $\kappa = \Sigma/\Sigma_c$ given by 
\begin{equation}
\kappa\left(R \right) = \frac{n}{2\sqrt{q}} \left[(R^2 + c^2)^{-\gamma} - (R^2 + t^2)^{-\gamma} \right].
\label{eq:kappa_dSPEMD}
\end{equation}
This corresponds to a 3D density
\begin{eqnarray}
 \rho(r) &=& \frac{\tilde n}{2\sqrt{q}}  \left[\left(r^2 + c^2\right)^{-\gamma_{\mathrm{3D}}/2} - \left(r^2 + t^2\right)^{-\gamma_{\mathrm{3D}}/2} \right]
\\ \nonumber
\mathrm{where}\quad\tilde n &=& n\frac{\Sigma_{\mathrm{c}} \Gamma\left[\gamma_{\mathrm{3D}}/2\right]}{\sqrt{\pi}\Gamma\left[\left(\gamma_{\mathrm{3D}}-1\right)/2\right]}.
\label{eq:kappa_dSPEMD_3D}
\end{eqnarray}
Here, the 2D slope $\gamma$ is related to the 3D slope via $\gamma=\left(\gamma_{\mathrm{3D}}-1\right)/2$.
$R^2 = x^2 + y^2/q^2$  denotes the projected, elliptical radius, where $q$ is the ellipticity, $c$ the core radius and $t$ the truncation radius.
The orientation of each mass component on the sky is described with a position angle $\phi$, measured in radians west to north.
The dSPEMD profile thus has a total of eight free parameters $(x,y,\phi,n,c,t,\gamma,q)$.

Since including eight parameters for each of our nine dSPEMD mass components requires  an optimization in a 72-dimensional very nonlinear parameter space, 
we have decided to fix those sets of parameters that are either not constrained by the data, or that do not influence the final quality of the model.
In particular, tests during which all parameters are allowed to vary, show that the slope of the cluster halo A and of the BCG do not affect the quality of the model and converge to a value close to $\gamma = 0.5$ (isothermal).
We believe that fixing the slope of A to isothermal, does not significantly affect our main results on total the central slope, since the latter is given by the sum of all mass components 
in the centre which are degenerate among each other. 
Similarly, we fix the ellipticity, the position angle and the centre position of  the BCG and of the galaxies G1 and G2 to those of their light distribution. The core radius for these galaxies is also fixed to a zero value, while
the normalization and the truncation radius are free parameters. For Galaxies G1 and G2 we additionally leave the slope of the dSPEMD as a free parameter.
This is different from previous analysis of galaxy truncations, where the slope is generally kept fixed at isothermal \citep{Suyu_2010_truncation,Donnarumma_2011,Eichner_2013_macsj1206}.
This simplifying assumption is often necessary to avoid degeneracies between the slope, the normalization and the truncation radius.
In most cases of \emph{isolated} galaxy-scale lensing there are not enough constraints on different radii to break those degeneracies since there is only
information at the scale of the Einstein radius. The unique case of \macs should provide enough constraints over a large range of radii to measure both the slope and the truncation radius for both the galaxies G1 and G2. 
We refer the reader to the forthcoming paper (Rau, Vegetti \& White, in preparation, hereafter \ppwo) for a detailed comparison of different mass models for G1 and G2. 
For the galaxies G3 and G4, we also fix their position to the centre of the light distribution, the slope to isothermal and the truncation radius to $1$ and $7 \asec$ respectively (as inferred from preliminary tests).
The galaxies B and S1 are modelled as round isothermal profiles and each have a free normalization parameter.
The galaxy group C is modelled as a cored elliptical mass distribution, where the normalization, the core radius, the position angle and the truncation radius are free to vary. 
In order to reduce the number of free parameters, we fixed the ellipticity of clump C to 0.6, as derived from an elliptical fit to the smoothed light distribution in the F555W filter. 
Clump C is, however, an approximation for the mass contribution a group of galaxies. 
Finally, the cluster halo A is not truncated. 

At larger radii we include {\em all galaxies with $\mathrm{I}_{814}< 20.5$} as isothermal dSPEMD with position, ellipticity, 
and rotation angle fixed to the best fitting parameters of their light distribution while the mass normalization, 
the core radius and the truncation radius are scaled with the galaxy luminosity via the following scaling relations
\begin{eqnarray}
 \sigma =& \sigma^*\left( \frac{L}{L^*}\right)^{1/4} \nonumber \\ 
 r_{\mathrm{core}} =& r_{\mathrm{core}}^*\left( \frac{L}{L^*}\right)^{1/2} \nonumber \\
 r_{\mathrm{cut}} =& r_{\mathrm{cut}}^*\left( \frac{L}{L^*}\right)^{1/2} 
\label{eq:gal_scaling_relations},
\end{eqnarray}
as previously done by \cite{Smith_2009}, \cite{Jullo_2007_clusters} and \cite{Eichner_2013_macsj1206}.  We adopt the best-fitting parameters for an
$L^*$ galaxy from \cite{Smith_2009}, $\sigma = 180 \mathrm{km/s}$, $r_{\mathrm{core}} = 0.2 \mathrm{kpc}$ and $r_{\mathrm{cut}}= 30 \mathrm{kpc}$. 

In the previous analysis of this cluster by \cite{Zitrin_2009} and  \cite{Smith_2009}, the masses of {\em all} galaxies, apart from the BCG, 
were tied to one fiducial galaxy via equation~\ref{eq:gal_scaling_relations}, 
while their position, ellipticity and position angle were fixed to those of their respective surface brightness distributions. 
In this work, instead, all five central galaxies and the three outer mass components are independently modelled.
This represents a major improvement in the model, since the details of the lensing reconstruction crucially depend on the central mass distribution.

Finally, the external shear is described by a shear strength $\gamma_{\mathrm{s}}$ and a position angle $\theta_\gamma$, both  free parameters of the mass model.

\section{Modelling Methods}
\label{sec:modelling_methods}
We model the mass distribution of \macs with two different methods. 
We first build a model using the lensed image positions and 
the constraints introduced in Sec.~\ref{sec:observations_and_constraints}, then,  as described in the next sections,
we refine this model with a hybrid model that includes both the position constraints plus the full image surface brightness distribution.  

\subsection{Modelling of the image positions}
\label{sec:modelling_pos}
The image position modelling is based on the optimization of the posterior 
\begin{equation}
P\left( \bm\eta | \bm d \right) \propto P(\bm d|\bm\eta)P(\bm\eta)\,,
\label{eq:pos_posterior}
\end{equation}
where 
\begin{equation}
\bm \eta = \left( x_i, y_i, \phi_i, n_i, \gamma_i, c_i, \epsilon_i, t_i\right) \quad i~\mathrm{in}~\left\lbrace 1,\dots,M \right\rbrace,
\end{equation}
is a vector containing the model parameters of the analytic mass distribution in Eq.~\eqref{eq:kappa_dSPEMD} and
$M$ is the number of mass components, $\kappa\left( R\right) = \sum_{i=1}^{M} \kappa_i  \left( R\right) $.
The data vector $\bm d$ contains all of the image position constraints.
The likelihood,
\begin{eqnarray}
 P(\bm d|\bm \eta) &=& \frac{1}{Z} \exp{\left(-\frac{1}{2}\sum_i^{N} 
  \delta \bm u_{i}^{T} \bm \mu_i^T \bm C_i^{-1} \bm \mu_i   \delta \bm u_i 
   \right)} \label{eq:sp_distance} \\ 
 \delta \bm u_{i} &=& \bm u_i^{\mathrm{obs}} - \bm u_i^{\mathrm{mod}} \nonumber
\label{eq:pos_likelihood}
\end{eqnarray}
is defined on the source plane, which is a good and fast approximation for the true distance on the image plane \citep[see for example][]{Halkola_2006,Suyu_2010_truncation}.
Here, the source plane positions are denoted as $\bm u_{i}$ and the sum is over the $N = 3 (4,5,7)$ images of \macs. 
The entries of the vectors $\bm u_{i}$ for the source positions and $\bm \mu_{i}$ for the magnification contain the different constraints for each image,
while the diagonal covariance matrix $\bm C_i$ contains the uncertainty $\sigma_i$ of the positional data. 
Z is the normalisation constant $1/(2 \pi \, \mathrm{det} \bm{C})$.

The speed and the power of this method depend on the number of source clumps that are identified as multiply lensed images and as constraints. 
Compared to the modelling of the full image brightness distribution described in the next section, the position modelling only uses a very small fraction of the available information resulting
in greater parameter uncertainties.
However, since the evaluation of equations \eqref{eq:pos_posterior} to \eqref{eq:sp_distance}
is very fast, we can use the position-based modelling to rule out a big portion of the full parameter space. 
In practice, we include this method also in the surface brightness modelling, by disfavouring  models that predict 
images offset from the observed true position that are separated by more than the size of the image 1.2.
This is a significant speed-up for the full image surface brightness modelling. 

\subsection{Hybrid modelling}
\label{sec:modelling_hybrid}
The full modelling of the image surface brightness information  
is based on the technique introduced by \cite{Suyu_2006} and
\cite{Vegetti_2009}. 
In the following, we summarize the most important aspects of this method.
At each step of the modelling, we find the best source surface brightness distribution $\bm s$ for an observed image
surface brightness distribution $\bm d$, source regularization 
strength $\lambda_{\mathrm{s}}$ and form $\bm R$. For each set of lens model parameters $\bm \eta$, we find the
source surface brightness distribution that maximizes the following probability density function
\begin{equation}
P\left(\bm s| \bm d, \bm \eta, \bm R \right) =   \frac{P\left(\bm d| \bm s, \bm \eta, \bm R \right) P\left( \bm s \right)}{P\left( \bm d | \lambda_s, \bm \eta, \bm R \right)}\,.
\label{eq:best_source_brightness}
\end{equation}
The likelihood 
\begin{equation}
 P\left(\bm d| \bm s, \bm \eta, \bm R \right) = \frac{1}{Z} \exp{ \left[-\frac{1}{2} \left(\bm M \bm s - \bm d  \right)^T \bm C_d^{-1} \left(\bm M \bm s - \bm d  \right) \right]}
\label{eq:Ed}
\end{equation}
is a measure of how well the model fits the data. $Z$ is the normalization, $\bm M$ is the lensing operator relating the source plane positions to the lens plane positions via the lens equation. $\bm M$ is calculated in each step
as a lensing matrix for a set of parameters $\bm \eta$. In Eq.~\eqref{eq:Ed}, $\bm C_d$ denotes the covariance of the observed images on the image plane.
For the modelling of the F555W data of \macs, we assume uncorrelated noise on the observed images with a Gaussian distribution with a rms of $\sigma = 0.0065$. 
We choose a quadratic prior for the source, $ P\left( \bm s \right) \propto \exp\left(-0.5 || \bm R \bm s||^2 \right)$, favouring a flat source brightness distribution.
In order to accommodate the very high dynamic range of the surface brightness distribution, we use a gradient source regularization.
Since the lensed image system 1 is very large, we only consider a subset of the image plane pixels (by a factor of 5) which are lensed back to the source plane. 
Those points on the source plane then define the base of a Delaunay triangulation which is used to interpolate to the full data set.
Implicitly finding the best source via Eq.~\eqref{eq:best_source_brightness} in each step, we then optimize for the best model parameters using the posterior
\begin{equation}
 P\left(\lambda_s, \bm \eta | \bm d, 	\bm R \right) = \frac{ P\left( \bm d | \lambda_s, \bm \eta, \bm R \right) P\left(\lambda_s, \bm \eta \right) }{ P\left( \bm d | \bm R \right) }.
\label{eq:posterior}
\end{equation}

In practice, when we model the surface brightness distribution of the largest image 1, we also include the positions of all other images as positional constraints. 
This is necessary, since none of the mass components B, C and S1 is constrained by the surface brightness information of image 1 alone. 
We do not include the full surface brightness information of the other image systems because those images are not very extended in the observations. 
We, therefore, perform a hybrid approach where the likelihood  $P\left( \bm d | \lambda_s, \bm \eta, \bm R \right)$ in Eq.~\eqref{eq:posterior}
is a multiplication of the position based likelihood in  Eq.~\eqref{eq:pos_likelihood} and the surface brightness based likelihood of Eq.~\eqref{eq:Ed} integrated over all 
possible source surface brightness distributions of S1.

\section{Results}
\label{sec:results}
First, we present the results from the image position modelling, then we refine this result using the full surface brightness information of the extended lensed images 
in a hybrid modelling approach in Sec.~\ref{sec:modelling_hybrid}. The best parameters from both modelling methods can be found in Tab.~\ref{tab:best_modelling_parameters}.

\begin{table}
  \caption[Inferred Modelled Lens parameters for position based and hybrid modelling]{
Inferred mass model parameters and $95 \%$ CL errors from the image position modelling and from the hybrid modelling of \macswo. 
Parameters $x,y,c,t$ are in $\mathrm{arcsec}$, and positions are defined relative to the BCG.
\label{tab:best_modelling_parameters}	
}
  \begin{center}
    \begin{tabular}{ c    c    c    c  }\hline
Mass component       & Parameter &       Hybrid model           &        Pos. model 		\\ \hline 
Halo  & $x_h$			&	  $ -1.67^{+0.011}_{-0.012} $        & 	$ -1.95^{+0.28}_{-0.26} $	\\
& $y_h$			&         $ -0.199^{+0.0109}_{-0.0096} $     & 	$ -0.53^{+0.27}_{-0.23} $ \\
& $\phi_h$		&         $ 0.605^{+0.0010}_{-0.0015} $    	&	$ 0.59^{+0.031}_{-0.023} $\\
& $n_h$			&         $ 15.76^{+0.080}_{-0.063} $        &	 	$ 15.0^{+0.79}_{-0.79} $\\
& $c_h$			&         $ 11.15^{+0.032}_{-0.050} $        & 	$ 12.04^{+0.91}_{-0.86} $\\
& $\epsilon_h$		&         $ 0.4457^{+0.0025}_{-0.0015} $	&		$ 0.491^{+0.023}_{-0.023} $ \\ \hline
BCG & $n_{\mathrm{BCG}}$		&        $ 1.0719^{+0.0043}_{-0.0052} $    &		 $ 1.015^{+0.075}_{-0.066} $\\
& $t_{\mathrm{BCG}}$		&        $ 20.9^{+1.2}_{-1.1} $            & 		$ 23.3^{+8.0}_{-8.0} $\\\hline
Galaxies &   $n_{g1}$		&       $ 0.2259^{+0.0081}_{-0.0063} $      & 		$ 0.20^{+0.16}_{-0.12} $\\
& $\gamma_{g1}$		&       $ 0.7558^{+0.0043}_{-0.0055} $ 	    &		$ 0.76^{+0.15}_{-0.18} $\\
& $t_{g1}$		&       $ 1.305^{+0.049}_{-0.064} $         & 		$ 4.9^{+4.0}_{-3.5} $\\
& $n_{g2}$		&       $ 0.3875^{+0.0035}_{-0.0027} $      & 		$ 0.135^{+0.111}_{-0.080} $\\
& $\gamma_{g2}$		&       $ 0.7274^{+0.0012}_{-0.0013} $ 	    &		$ 0.910^{+0.081}_{-0.104} $\\
& $t_{g2}$		&       $ 6.95^{+0.41}_{-0.46} $            & 		$ 5.7^{+6.2}_{-4.7} $\\
& $n_{g3}$			&       $ 0.3254^{+0.0022}_{-0.0020} $      & 		$ 0.348^{+0.070}_{-0.062} $\\
& $n_{g4}$		&       $ 0.2596^{+0.0027}_{-0.0021} $      & 		$ 0.144^{+0.069}_{-0.084} $\\\hline
B and C &$n_{B}$			&       $ 3.630^{+0.081}_{-0.042} $         & 		$ 3.20^{+0.18}_{-0.23} $\\
& $\phi_C$		&       $ 2.250^{+0.025}_{-0.024} $         & 		$ 2.211^{+0.095}_{-0.116} $\\
& $n_{C}$			&       $ 90.7^{+2.3}_{-3.5} $              & 		$ 132^{+14}_{-16} $\\
& $c_{C}$			&       $ 62.1^{+1.4}_{-1.6} $              & 		$ 51.6^{+4.5}_{-3.0} $\\
& $t_{C}$			&       $ 82.2^{+1.9}_{-2.3} $              & 		$ 79.1^{+6.0}_{-4.3} $\\ \hline
Source & $x_{S1}$		&	$	-1.59$		    &			$3.77$		\\
& $y_{S1}$		&		$4.92$			    &		$13.48$			\\
& $n_{S1}$		&       $ 1.68^{+0.26}_{-0.23} $            & 		$ 0.058^{+0.093}_{-0.036} $\\\hline
Shear & $|\gamma|$		&       $ 0.0262^{+0.0011}_{-0.0014} $      &		$ 0.0398^{+0.0075}_{-0.0085} $\\
& $\phi_\gamma$		&       $ 2.327^{+0.012}_{-0.015} $    	    &		$ 1.290^{+0.134}_{-0.097} $\\\hline
    \end{tabular}
  \end{center}
\end{table}

\subsection{Results from the image position modelling}
\label{sec:results_modelling_pos}
\begin{figure}
\begin{center}
\includegraphics[trim=0mm 0mm 0mm 0mm,clip,width=1.0\columnwidth]{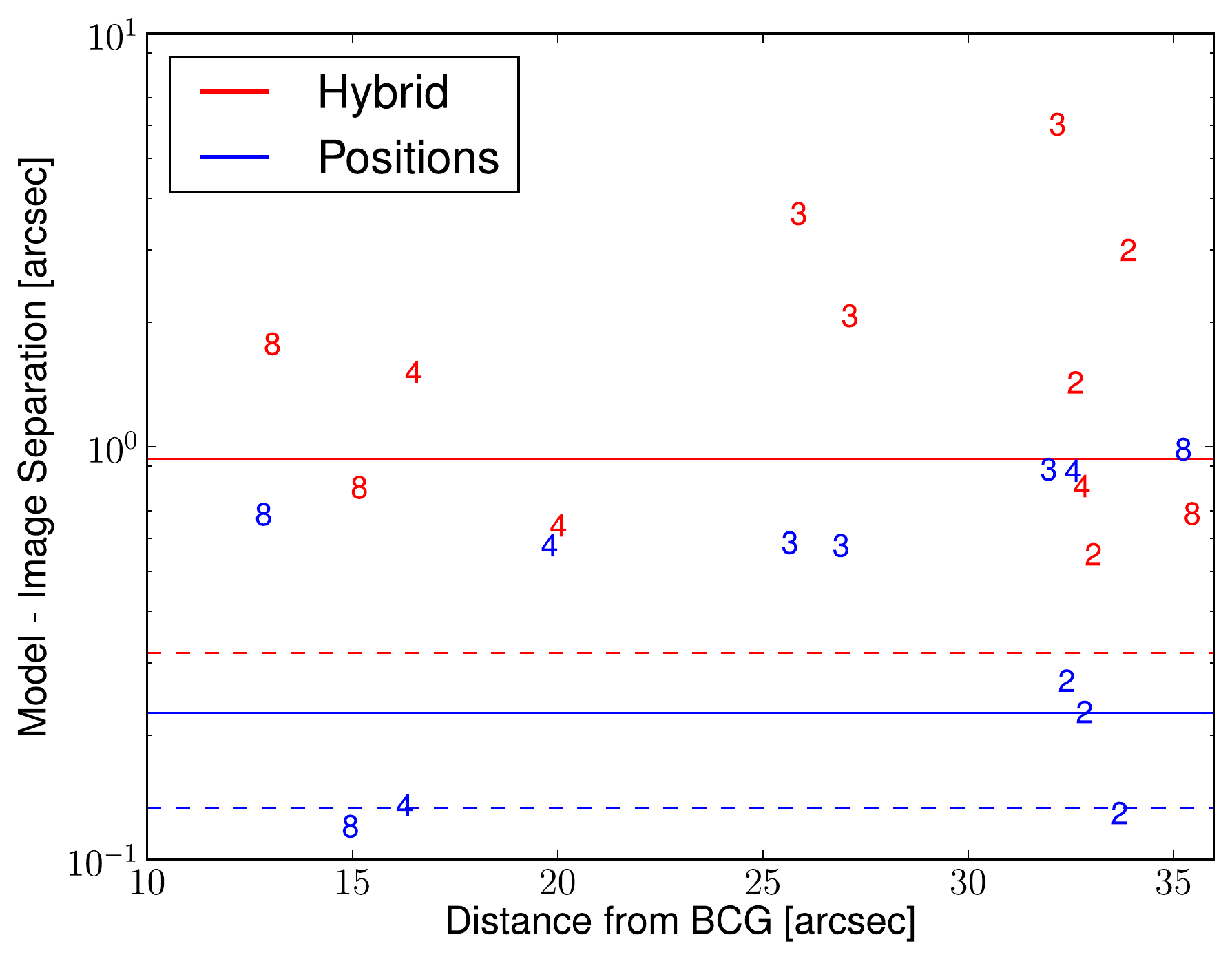}\\
\includegraphics[trim=0mm 0mm 0mm 0mm,clip,width=1.0\columnwidth]{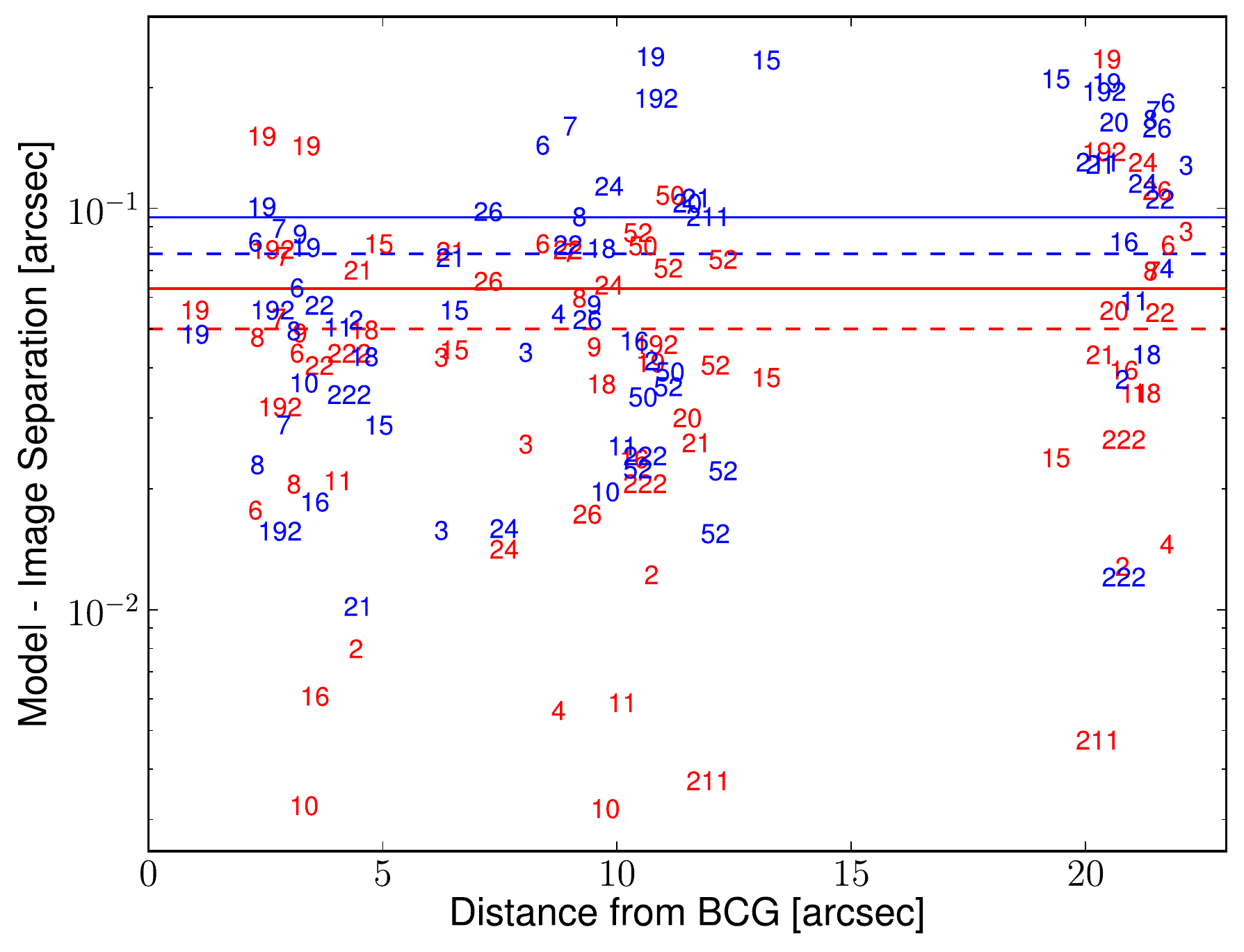}
\end{center}
\caption[Model-Image Separation with Distance from the BCG]{
Separation between the inferred and the observed image position as a function of increasing distance from the BCG.
Top panel: outer constraints. Bottom panel: constraints in image system 1. See Figs.~\ref{fig:constraints_four} and \ref{fig:HST_cluster} for the notation.
Blue numbers show the results from the position based modelling while red numbers show the results from the hybrid modelling technique. 
Solid lines are the root mean square, dashed lines are the mean for all constraints (top panel) and the constraints in image 1 (lower panel). 
\label{fig:sp_pos}
}
\end{figure}
Fig.~\ref{fig:sp_pos} shows in blue all the constraints that we have used in the position based modelling method as a function of their respective distance to the BCG for our best model. 
The y-axis is the goodness-of-fit, that is, the distance between the best modelled image position and the observed image position in $\mathrm{arcseconds}$ on the image plane.
The outer constraints are shown in the top panel of Fig.~\ref{fig:sp_pos} while the lower panel focuses on the constraints in image 1 (cf. Fig.~\ref{fig:constraints_four}).
Blue solid and dashed lines are the rms and the mean values, respectively. The values are $\mathrm{rms}=0.23 \asec$ and $\mathrm{mean}=0.14\asec$ for all constraints in the top panel
and $\mathrm{rms}=0.096 \asec$ and $\mathrm{mean} = 0.08\asec$ for the constraints in image 1 only in the lower panel.

We use a positional error in the measurements of the positions of the images in Eq.~\eqref{eq:pos_likelihood} of $\sigma = 0.2 \asec$ for the central constraints in image 1 at redshift $z=1.49$.
For the constraints in images 2, 3, 4 and 8, we use a higher uncertainty, $\sigma=0.78 \asec$, since they are at higher redshifts and 
at least one of the multiple images of those systems is at a greater radius.
This higher uncertainty leads to a radial dependence of the fit quality in Fig.~\ref{fig:sp_pos}.

\subsection{Results from the hybrid modelling}
\label{sec:results_modelling_brightness}
\begin{figure}
\begin{center}
\includegraphics[width=1.0\columnwidth]{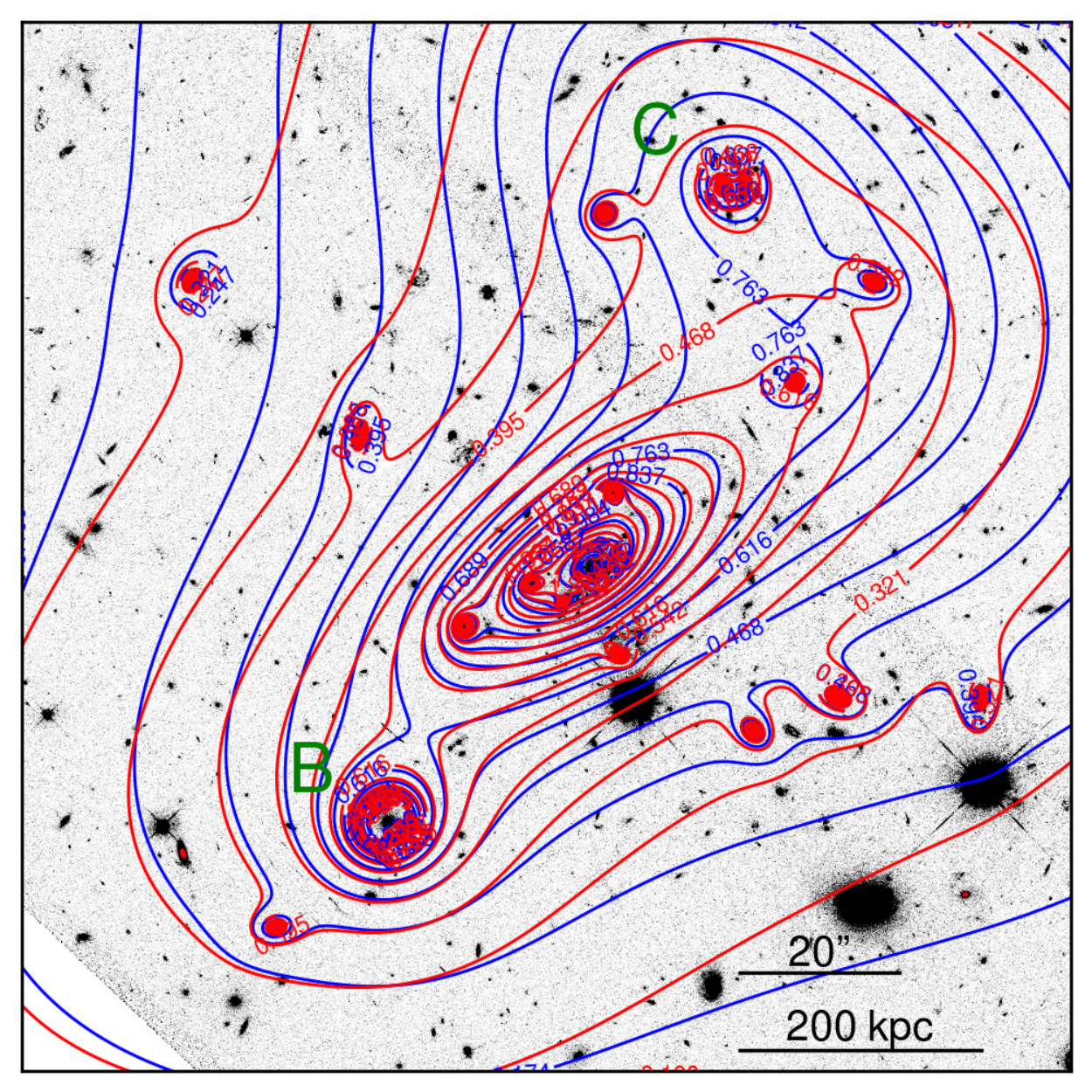}
\end{center}
\caption[Contours of the mass distribution]{
Contours of the cluster mass distribution at redshift 1.49, including scaled cluster galaxies and the two additional mass components B and C. 
Blue contours show the results from the position-based modelling, while red contours from the hybrid modelling.
The grey background shows the CLASH observation of the cluster in the F555W filter.
\label{fig:contours_mass_light}
}
\end{figure}
The hybrid modelling technique includes all the positional constraints of Sec.~\ref{sec:modelling_hybrid}, plus the full surface brightness distribution of image 1.
First, we evaluate the quality of the results of this modelling in terms of the image positions. 

The red numbers in Fig.~\ref{fig:sp_pos} indicate the distance between modelled and observed image positions on the image plane. 
The rms distances of the separation of all (solid red line in the top panel of Fig.~\ref{fig:sp_pos}) and central constraints (solid red line lower panel) are $0.94$ and $0.063\asec$, respectively, 
the means are $0.32$ and $0.05$ respectively (red dashed lines).

Due to the increased number of constraints, the hybrid modelling technique puts more emphasis on the accurate reconstruction of the surface brightness distribution
of the central image 1. Therefore, the respective rms is improved to a level comparable to the resolution limit of the CLASH data. Instead, the outer constraints are weighted less and consequently
the image positions of the other images are reproduced less perfectly. 
As a consequence, the reconstruction of all multiple images is worse in terms of the total rms.
However, the model based only on positions from Sec.~\ref{sec:results_modelling_pos} performs worse for the central image:
the solution is usable as a starting point for the hybrid model, but it does not reproduce the image surface brightness distribution in detail.

Fig.~\ref{fig:contours_mass_light} shows the contours of the scaled surface mass density overlaid over a grey background image of the F555W ACS observation for both our models.
While the centre of the cluster mass distribution of both models is very similar, there are differences in the modelling of the mass component C in the upper right. 
This is because this mass component is mainly constrained by the upper multiple images of systems 3 and 4 with a relatively high positional uncertainty. 
Those constraints therefore do not provide enough information to constrain the mass component C tightly.
Figure~\ref{fig:contours_mass_light} shows the integrated 2D line-of-sight $\kappa$ contours for the main source at redshift $z=1.49$.
This does {\em not} include the mass distribution of the source S1 itself. 

We find that the recovered position of the mass component S1 is different between the two modelling approaches. 
We believe this change to be related to the difference in the inferred mass of the component C. 
In fact, the change in the deflection angle of C from one model to the other is essentially equivalent to the measured shift  in the position of S1, 
that is $\Delta\alpha_{\rm{C}}\sim-\Delta r_{\rm{S1}}\sim10.1\asec$.

\begin{figure*}
\begin{center}
\includegraphics[width=2.0\columnwidth]{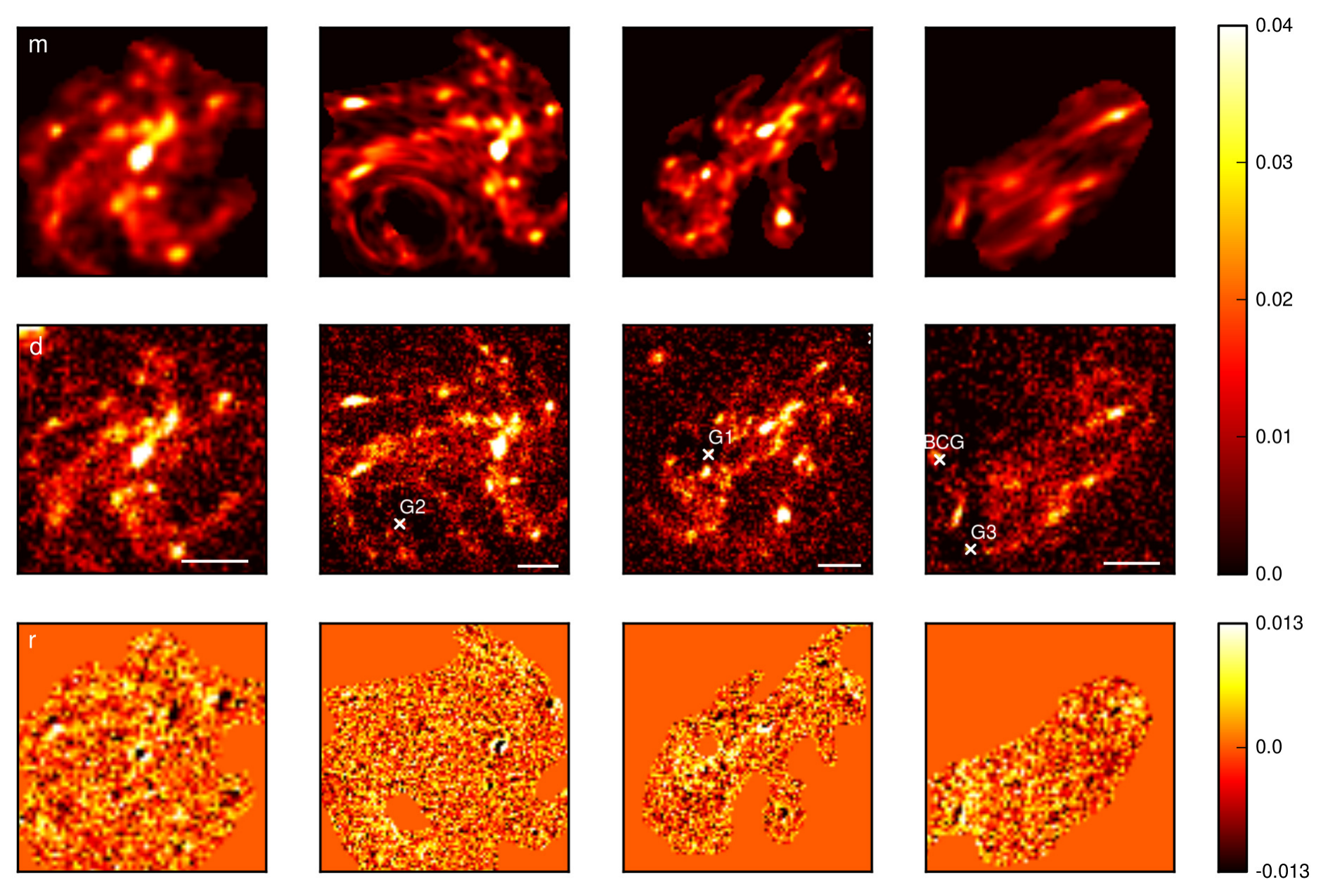}
\end{center}
\caption[Surface brightness modelling results]{
Results from the surface brightness modelling. Upper row is the model $\bm m$, middle row the observed data $\bm d$, lower row the residual $\bm r = \bm d - \bm m$. 
Columns from left to right are the main images 1.1, 1.2, 1.3.1 and 1.3.2.
The colour scale of the residuals is based on twice the rms of the pixel noise on the image plane. 
All features of the images are reproduced by our model down to the noise level except where the resolution of the data $\bm d$ is insufficient to capture
the high dynamic range at the edges of the very bright source clumps. In each column, the scale of 1 arcsec is indicated by a white line.
\label{fig:res_brightness}
}
\end{figure*}
\begin{figure}
\begin{center}
\includegraphics[width=1.0\columnwidth]{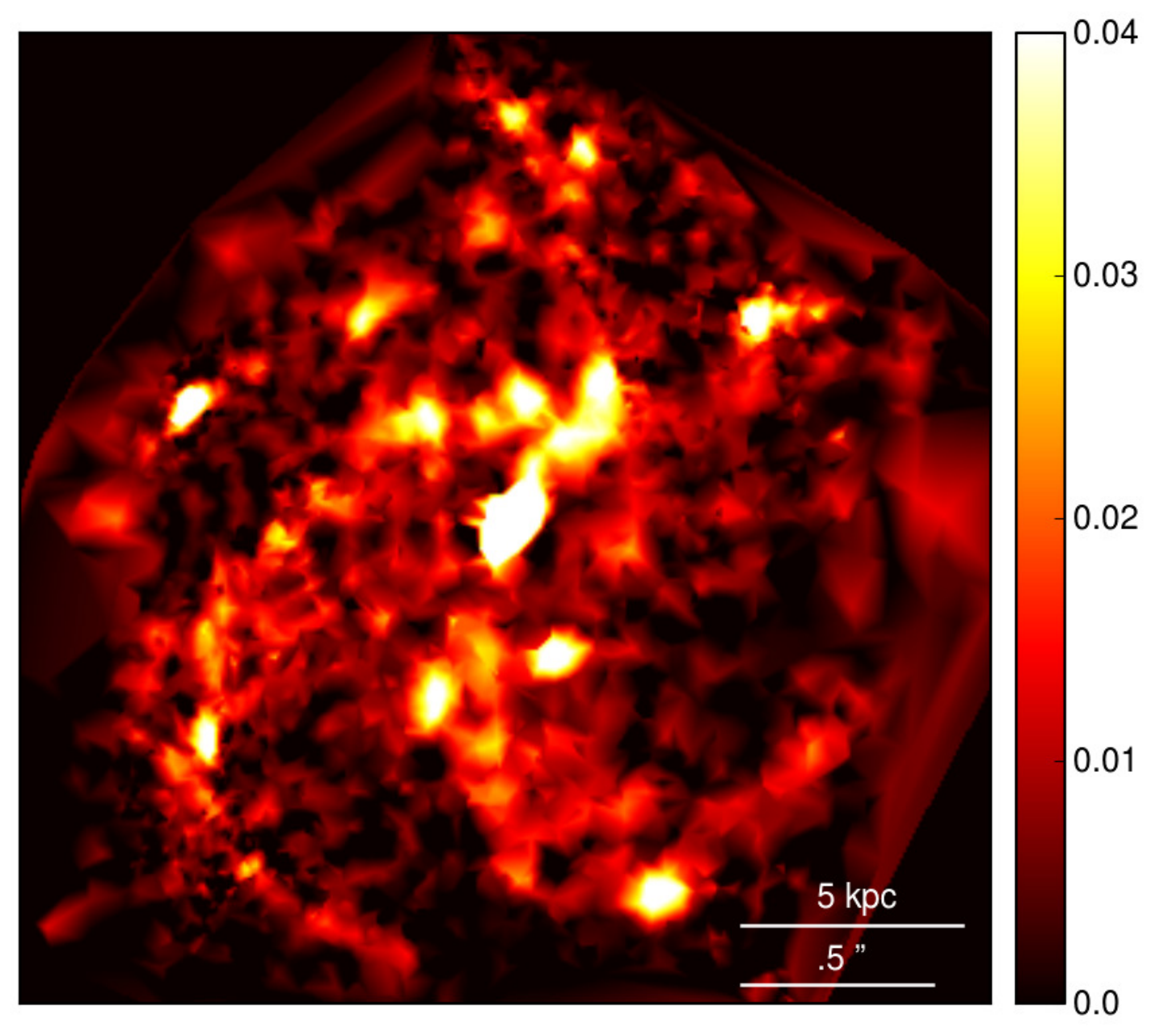}
\end{center}
\caption[Reconstructed source brightness distribution]{
Reconstructed source brightness distribution. Colour scale is the same as in Fig.~\ref{fig:res_brightness}. 
\label{fig:source_brightness}
}
\end{figure}
In the following, we present the reconstruction of the surface brightness distribution of image 1.
Fig.~\ref{fig:res_brightness} shows our best model of the lensed images. 
Each column contains one of the four main images of system 1: 1.1, 1.2, 1.3.1 and 1.3.2. 
In rows from top to bottom are the modelled images $\bm m$, the observed data $\bm d$ and the residual $\bm r = \bm m - \bm d$.
In the first column, image 1.1 is almost undistorted and closely resembles the original source surface brightness distribution. 
As a comparison, the best model of the source is shown in Fig.~\ref{fig:source_brightness}. 

The second image in the second column is distorted more significantly. 
The aforementioned Einstein ring in image 1.2 is very accurately reproduced by our model. 
Note also the additional multiple images $15$ and $21$ (for nomenclature see Fig.~\ref{fig:constraints_four})
in image 1.3.1 and all of the morphological details of image 1.3.2 are perfectly reproduced to the noise limit of the CLASH observations. 
In order to quantify the residuals, we define a pixelized rms distance on the image plane $\mathrm{rms}=\sqrt{\sum_i \left( d_i - m_i \right)^2/N}$. 
Our best model has a mean square distance of $1.14~\sigma$ averaged over the whole image 1, where $\sigma$ is the background noise in the CLASH observations.
There are minimal increased residuals at the positions of the brightest source clumps $2, 3, 15$ and $19$. 
Those are a consequence of the constant strength of the regularization used throughout our modelling. 
The surface brightness distribution of the source in Fig.~\ref{fig:source_brightness} is very non uniform and at the edges of the brightest source clumps the gradient increases significantly, approaching infinity. 
This simply means that in those regions the resolution provided by CLASH is insufficient to capture the brightness distribution in details. 
In this case, any form of regularization enforces a smoothing of those regions.

\citet{Yuan_2011_metallicity_gradient} have reported over the extent of image 1.2 a constant stretching of about 5. 
Over the same region, we find that the magnification changes from roughly 4 to 20 in the y-direction and reaches almost infinity in the x-direction 
for the same orientation chosen by \citet{Yuan_2011_metallicity_gradient}. 
This discrepancy is related to the fact that \citet{Yuan_2011_metallicity_gradient} have used the lens model by \citet{Smith_2009}, 
which does not include the contribution from the galaxy G2. 
In our case, the presence of this galaxy significantly changes the magnification in this region as expected from the observed Einstein ring around galaxy G2.

It is evident from the comparison of the hybrid modelling with the position modelling in 
Tab.~\ref{tab:best_modelling_parameters} 
that the best parameters do not always agree within the error bars. 
This is not unexpected, since by adding the surface brightness constraints to the positional constraints, we are effectively changing the data basis for the reconstruction. 
Therefore, parameter regions allowed during the modelling with positional constraints are discarded when the additional surface brightness information is used. 
Since the surface brightness contains at least two orders of magnitude more constraints and we are optimizing in a 24 parameter space, 
the best parameters from the hybrid modelling might be in a previously low-probability region.

The large central images of the spiral source galaxy constrain the central 2D logarithmic slope $\gamma_{\mathrm{2D}} = \partial \log \left(\kappa \right)/ \partial \log  R$ 
of the two cluster member galaxies G1 and G2. At a radial distance corresponding to their respective Einstein radii in the cluster of $1$ and $1.3 \asec$ , we find a slope of 
-2.51 and -1.63 respectively (spherical deprojected -3.2 and -2.6 in 3D). The mass profile of those galaxies is therefore steeper than that of  {\em isolated} field 
galaxies. For example \cite{SLACSIII_2006} and \cite{SLACSIV_2007} found in the SLACS sample a 3D slope that is consistent with isothermal $\rho \sim  r^{-2}$
in a range from $3$ to $300 \mathrm{kpc/h}$. 
Since G1 and G2 are in the dense environment of a galaxy cluster the tidal stripping of galaxy mass during the merging with the cluster is a plausible 
hypotheses for this discrepancy \citep{Merrit_1983,Merrit_1984,Merrit_1985,Ghigna_1998,Diemand_2007,Gao_Phoenix}. 
For a detailed analysis and the comparison with different galaxy mass profiles, we refer the reader to \pp.

\subsection{Parameter Degeneracies}
\label{sec:Parameter_Degeneracies}
\begin{figure*}
\begin{center}
\includegraphics[width=2.0\columnwidth]{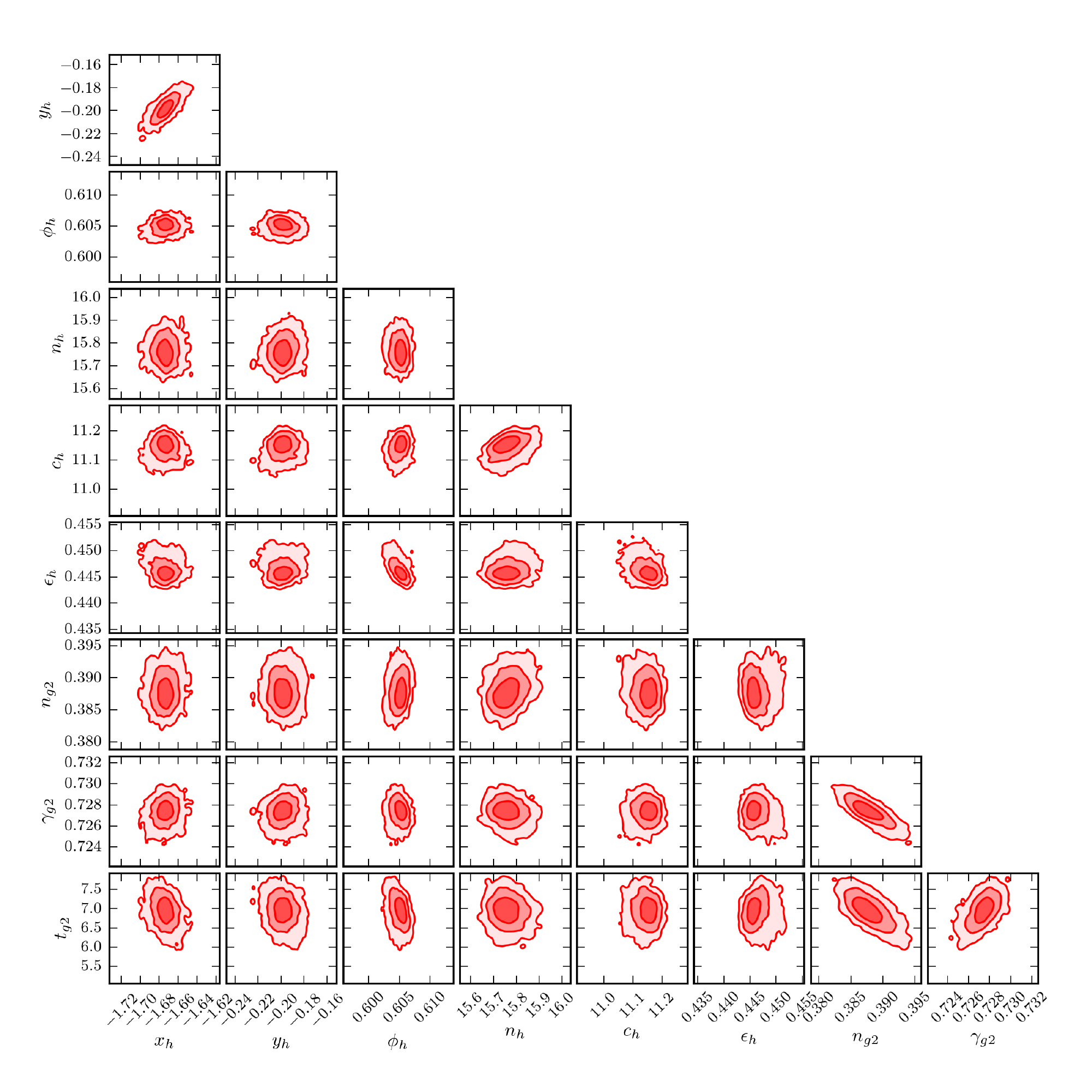}
\end{center}
\caption[Correlations between modelling parameters]{
2D marginalized posteriors for a selection of eight parameters from the hybrid modelling.
The inclusion of the surface brightness information, allows us to tightly constrain all model parameters.
The parameters $x,y,c,t$ are in $\mathrm{arcsec}$, and $x$ and $y$ are defined relative to the BCG.
\label{fig:corr_br_pos}
}
\end{figure*}
Our best model has 24 free parameters, six for the main DM halo, ten for the cluster galaxies, six for mass components farther from the BCG and two for the external shear.  
In order to quantify the degeneracies among these parameters, we perform an exploration of the parameter space
with the publicly available library {\sc MultiNest} \citep{Multinest_I, Multinest_II, Multinest_III}.
As an example, we show in Fig.~\ref{fig:corr_br_pos} the degeneracies for a selection of eight parameters, including 
all the parameters for the central halo A and all the parameters for the cluster galaxy G2.
Note that we choose a flat prior within the range $0.065\asec < t_{\mathrm{G2}} <  13 \asec$
for the truncation of the substructure galaxy G2 spanning the whole range to the central images 1.3.1 and 1.3.2 (cf. Fig.~\ref{fig:HST_cluster}).
In Fig.~\ref{fig:corr_br_pos}, we show the 68, 95 and 99.7 confidence levels (CL) from the hybrid modelling.
The simpler position-based modelling works well for most parameters. 
However, the surface brightness modelling provides a huge improvement on the modelling accuracy and, therefore, has also a huge effect on the size of the confidence regions.

The most noticeable example is the truncation radius of the substructure galaxy, $t_{\mathrm{G2}}$ in the bottom row.
Similarly to what was found by \cite{Suyu_2010_truncation}, 
the position modelling provides almost no constraints on the truncation radius ($t_{\mathrm{G2, pos}} =  5.7^{+6.2}_{-4.7}\asec$).
Therefore, the addition of the surface brightness distribution is crucial in order to constrain the truncation radius. 
Tight constraints on all galaxy parameters such as the normalization, 
the slope and the truncation radius allow us to constrain the total galaxy profile and therefore its size in detail. 
We find similar results for the modelling of galaxy G1. 
We have to keep in mind, however, that the confidence limits in Fig.~\ref{fig:corr_br_pos} do \emph{not} include systematic errors. 
Even though we get a very well constrained result from the hybrid modelling, these results
might be biased by the assumption made on the analytic form of the parametrization for the mass distribution \citep[see e.g.][]{schneider13}.
We refer to paper~\pp for a more thorough investigation of these effects.

\section{Central Slope of the Total Mass Distribution}
\label{sec:Central Slope}

\begin{figure}
\begin{center}
\includegraphics[width=1.0\columnwidth]{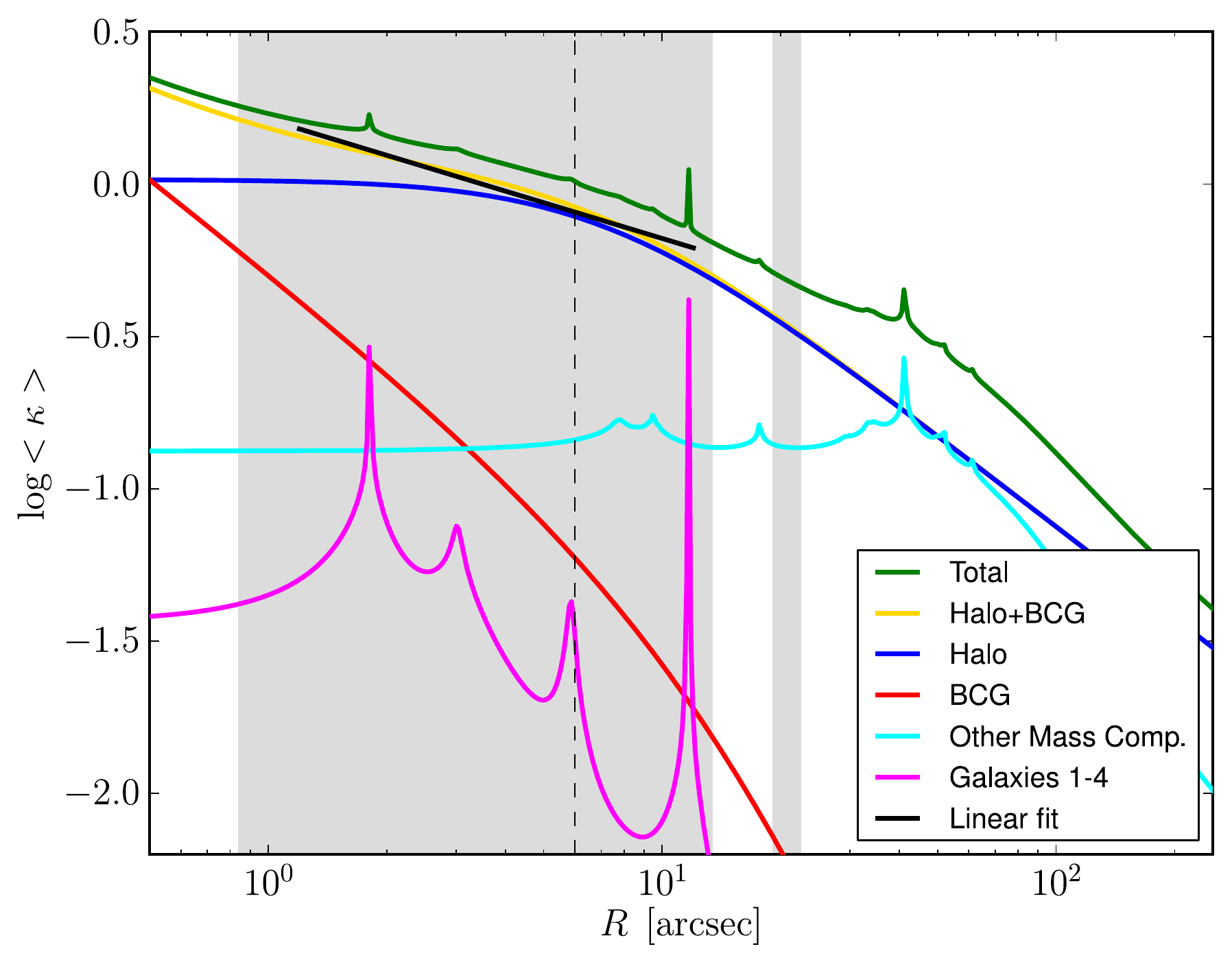}	
\end{center}
\caption[Radial mass density distribution and mass density slope]{
Radial distribution of the projected mass density $\kappa\left( R \right)$ as a function of 2D radius. 
The constituents of the total mass profile (green) are the main DM halo (blue), the BCG total mass (red) and central cluster galaxies (magenta), 
outer mass components (cyan). The DM halo has a large core of $\sim 12\asec$. Grey shaded areas show the radial coverage of the main image system 1.
The black line is a linear fit in log-log space to DM halo+BCG mass density, for details see text. Vertical dashed line is the virial 
radius of the BCG. 
\label{fig:kappa_radial}
}
\end{figure}
Figure~\ref{fig:kappa_radial} shows the radial distribution of the projected mass density $\left \langle \kappa \right \rangle$ for our best model,
where the average is over circles centred on the BCG.
The grey shaded areas indicate the radial coverage of the main images 1.1, 1.2 and 1.3. 
Note, that the radial extent of the images 1.1 and 1.2 overlaps.
The three main images cover almost the entire range from $0.8$ to $22\asec$ $(5.3$ to $145~\mathrm{kpc})$. 
Figure~\ref{fig:kappa_radial} differentiates the contributions of the main dark matter halo, the BCG, the galaxies G1 to G4 and the mass components B and C and the scaled galaxies.
The dark matter component of halo A has a large core and is flat out to $\sim 10 \asec$ $(\sim 66~\mathrm{kpc})$.
The central cusp of the total mass distribution for $R < 5  \asec$ $(33~\mathrm{kpc})$ gets increasingly dominated by the profile of the BCG towards the centre. 
Due to the particular mass distribution of \macswo, the contribution of the other cluster galaxies G1 to G4 to the innermost radial mass distribution is also non negligible 
(see below).

For the total mass distribution, we measure a 2D logarithmic slope of $\gamma_{\mathrm{2D}} = \partial \log{<\kappa>}/\partial \log{R}$, 
in the range between $1.2$ and $12~\mathrm{arcsec}$  ($8$ to $80~\mathrm{kpc}$), 
this corresponds to $0.2$ to $2~R_e$ for a BCG half light radius of $\sim 6\asec$. This is the same range probed by \citet{Newman_2013a}.
Note, that our surface brightness reconstruction of the cluster main image system covers the whole range, so the reconstructed mass distribution is very well constrained.
The straight black line in the upper panel of Fig.~\ref{fig:kappa_radial} is a linear fit in $\log{\kappa} - \log{R}$ space (equally spaced in $\log{R}$) 
to the DM halo+BCG mass density distribution. 
We find a mass density slope for the DM halo+BCG mass components of $\gamma_{\mathrm{2D}} \approx  -0.39$. 
The green line in the top panel of Fig.~\ref{fig:kappa_radial} additionally includes the mass distributions of the four central galaxies G1 to G4, all scaled galaxies and the mass components
B and C. We find that in the radial range from $1.2$ to $12 \asec$ these components contribute considerably to the total mass distribution.
The total 2D density slope $\gamma_{\mathrm{tot,2D}}  \approx -0.33$ is shallower than that of the DM halo + BCG alone.
We conclude, therefore, that cluster galaxies located very close to the cluster centre, can have a significant contribution to the total potential in which the stars of the BCG form. 

It is therefore important to accurately model and to include the mass distribution of cluster member galaxies
in order to measure the total mass distribution.
In order to estimate the scatter in the slope measurement due to the galaxy subhalo population, we exclude the galaxies G1 to G4.
If for example the two more distant galaxies were excluded, G1 and G2, the averaged total density slope would increase to $\gamma_{\mathrm{tot}}  \approx -0.37$.
Equivalently, excluding the innermost galaxies, G3 and G4, results in a decrease of the total density slope, $\gamma_{\mathrm{tot}}  \approx -0.32$.

Our inferred 2D total logarithmic mass density slope of $\gamma_{\mathrm{tot,2D}} \approx  -0.33$ is slightly steeper than the values from the two independent measurements 
by \cite{Smith_2009} and \cite{Zitrin_2009}, $-0.31$ and $-0.27$, respectively (from their figs. 4 and 5).   
We note that they do not include the full surface brightness distribution of the main system 1
that covers the centre of the cluster and only use $\sim 10$ constraints in each of the images 1.1, 1.2 and 1.3.
As a comparison, our best model based on twice as many image positions as previously used has a DM halo+BCG logarithmic mass density slope 
of $\gamma_{\mathrm{2D}} \approx  -0.35$ and
a total slope including all mass components of  $\gamma_{\mathrm{2D}} \approx  -0.28$.
This indicates that the reconstruction using the full surface brightness information by the hybrid modelling technique is important for the correct reconstruction of the central
mass density slope.

Using spherical deprojection, our model predicts a spherically averaged 3D logarithmic slope of $\gamma_{\mathrm{3D}} \approx  -1.13$ for the DM halo+BCG components.
\citet{Newman_2013a} have measured the slope of the mass distribution of the DM halo + BCG mass components for seven massive clusters and 
have found a mean central spherically averaged logarithmic slope of $<\gamma> = -1.16$ 
with intrinsic scatter $\sigma_{\gamma} = 0.13~(68 \% \mathrm{CL})$.
Our density slope is therefore within $0.2~\sigma_{\gamma}$ the value derived by \citet{Newman_2013a}.
A certain difference is to be expected considering the differences between \macs and the cluster sample from \citet{Newman_2013a}:
\macs with a mass of $M_{200} \sim 5\ten{14}\Msun$ is at the lower end of the mass range $0.4<M_{200}/(10^{15} \Msun) < 2$.
The cluster is at higher redshift $z = 0.54$ compared to their sample ($0.2<z<0.3$) and \citet{Newman_2013a} choose relaxed clusters. 
In contrast, there are several indications that \macs is not yet fully relaxed:
(a) the offset between the BCG and the cluster centre, (b) the high ellipticity of the cluster mass distribution, 
(c) close-by massive galaxies or groups of galaxies, B and C in Fig.~\ref{fig:HST_cluster}
and (d) non relaxed X-ray emission \citep[see for example fig.~3 in][]{Smith_2009}.

Without measurements of the BCG velocity dispersion for \macswo, it is not straightforward to separate the DM from the stellar mass content of the BCG. 
However, \cite{Zitrin_2009} reported a mass of $\sim 1\ten{12} \Msun$ for their BCG model component within the low surface brightness wings $(\lesssim 30\mathrm{kpc})$, 
which is identical to what we find here. They claim that the mass to light radius of $M/L_B = 4.5$ can be explained by the stellar content of a 
single burst stellar population formed at redshift $z = 3$ and a mean half solar metallicity. 
Under this assumption, the central logarithmic slope of the dark matter content of \macswo, the blue line in Fig.~\ref{fig:kappa_radial} is shallower than
the NFW profile. In fact, the central DM density is flat for $r<20~\mathrm{kpc}$.
This suggests that in the case of \macs the baryons at the cluster centre have flattened the dark matter distribution with respect to what is expected 
from purely dark matter simulations.

\section{Conclusions}
We have presented a new and detailed model for the centre of the galaxy cluster \macswo. In particular, we have identified more 
than twice as many constraints as previously used.  We have also used a multiple lens plane algorithm  in order to properly include the lensing 
contribution of the mass associated with the source S1. Finally, with a hybrid modelling approach, we have performed the first detailed reconstruction of 
the surface brightness distribution of the system 1. Our results can be summarized as follows:

 \begin{enumerate}
 \item We have recovered the surface brightness distribution of system 1 with a precision close to the noise level of the HST CLASH observations.
 \item Thanks to the hybrid modelling approach, we have derived posterior probability density distributions of the main model parameters that are significantly tighter than those
         derived with the simpler position modelling. 
\item Thanks to the new constraints, we have constrained three important details of the mass distribution: 
        the individual mass distributions of the two cluster galaxies G1 and G2, and the  
        total mass distribution of the cluster at the innermost radii. 
 \item  We  have recovered the 2D logarithmic slopes $\gamma_{\mathrm{2D}} = \partial \log \left(\kappa \right)/ \partial \log  R  \approx -2.51$ and $ \approx -1.63$ for galaxies G1 and G2 measured 
	  at a distance corresponding to their respective Einstein radius in the cluster.
  \item Our mass model suggests a large $(\sim 12~\mathrm{arcsec})$ core in the cluster DM distribution and that the total mass profile at the very centre of the cluster is dominated by the BCG. 
	We have found a central logarithmic slope of the 2D mass distribution between $1.2$ and $12~\mathrm{arcsec}$ of 
	$\gamma_{\mathrm{2D}} = \partial \log \left(\kappa \right)/ \partial \log  R  \approx -0.39$ for the DM halo+BCG mass components
         and  $ \approx -0.33$ when the other central galaxies and cluster members are included.
	
\end{enumerate}

\section{ACKNOWLEDGEMENTS}

SW is supported by Advanced Grant 246797 `GALFORMOD' from the European Research Council.

\bibliographystyle{mn2e}

\bibliography{manuscript}

 \end{document}